\documentclass[10pt,pra,aps,twocolumn,floatfix,nofootinbib]{revtex4-1}

\usepackage[utf8]{inputenc}
\usepackage[T1]{fontenc}
\usepackage{lmodern} 
\usepackage[ngerman,english]{babel}
\usepackage{graphicx}
\usepackage{microtype}
\usepackage{mathtools}
\usepackage{bbold}
\usepackage{xcolor}
\usepackage{hyperref}
\hypersetup{colorlinks,citecolor=black,filecolor=black,linkcolor=blue,urlcolor=blue}

\newcommand{\GeV}{\,\mathrm{GeV}}
\newcommand{\TeV}{\,\mathrm{TeV}}
\renewcommand{\Im}[1]{\,\mathrm{Im}\!\left[#1\right]} 
\renewcommand{\Re}[1]{\,\mathrm{Re}\!\left[#1\right]} 
\newcommand{\lnx}[1]{\,\mathrm{ln}\!\left(#1\right)} 
\newcommand{\diag}[1]{\,\mathrm{diag}\!\left(#1\right)} 
\newcommand{\overbar}[1]{\mkern 1.5mu\overline{\mkern-1.5mu#1\mkern-1.5mu}\mkern 1.5mu}
\newcommand*{\rom}[1]{\text{\expandafter \MakeUppercase{\romannumeral #1}}}

\begin{document}

\title{Low-Scale Leptogenesis in the Scotogenic Neutrino Mass Model}
\author{Thomas Hugle} \email{thomas.hugle@mpi-hd.mpg.de}
\author{Moritz Platscher} \email{moritz.platscher@mpi-hd.mpg.de}
\author{Kai Schmitz} \email{kai.schmitz@mpi-hd.mpg.de}
\affiliation{Max-Planck-Institut für Kernphysik (MPIK), Saupfercheckweg 1, 69117 Heidelberg, Germany}


\begin{abstract}
The scotogenic model proposed by Ernest Ma represents an attractive and minimal example for the generation of small Standard Model neutrino masses via radiative corrections in the dark matter sector. In this paper, we demonstrate that, in addition to neutrino masses and dark matter, the scotogenic model also allows to explain the baryon asymmetry of the Universe via low-scale leptogenesis. First, we consider the case of two right-handed neutrinos (RHNs) $N_{1,2}$, for which we provide an analytical argument why it is impossible to push the RHN mass scale below $M_1^\text{min} \sim 10^{10} \GeV$, which is identical to the value in standard thermal leptogenesis in the type-I seesaw scenario with the same washout strength. Then, we present a detailed study of the three-RHN case based on both an analytical and a numerical analysis. In the case of three RHNs, we obtain a lower bound on the $N_1$ mass of around $10\,\text{TeV}$. Remarkably enough, successful low-scale leptogenesis can be achieved without any degeneracy in the RHN mass spectrum. The only necessary condition is a suppression in the $N_1$ Yukawa couplings, which results in suppressed washout and a small active neutrino mass of around $10^{-12}\,\text{eV}$. This leads to the fascinating realization that low-scale leptogenesis in the scotogenic model can be tested in experiments that aim at measuring the absolute neutrino mass scale.
\end{abstract}


\maketitle

\section{Introduction}\label{sec_intro}

The baryon asymmetry of the Universe (BAU)\,---\,conventionally quantified in terms of the cosmic baryon-to-photon ratio $\eta_B^{\rm obs} \simeq 6.1 \,\cdot\, 10^{-10}$~\cite{Patrignani:2016xqp,Ade:2015xua}\,---\,cannot be explained within the Standard Model (SM) of particle physics. It, thus, provides compelling evidence for the existence of new physics beyond the SM. An attractive possibility to dynamically generate the BAU in the early Universe is baryogenesis via leptogenesis~\cite{Fukugita:1986hr}. In its standard formulation, leptogenesis is closely related to the type-I seesaw mechanism~\cite{Minkowski:1977sc,Yanagida:1979as,Yanagida:1980xy,GellMann:1980vs,Mohapatra:1979ia} that aims at explaining the small SM neutrino masses by introducing two or more sterile right-handed neutrinos (RHNs) $N_i$ with large Majorana masses $M_i$. In standard thermal leptogenesis, the heavy RHNs are produced through scatterings in the thermal bath, before their $CP$-violating out-of-equilibrium decays generate a primordial lepton asymmetry. This lepton asymmetry is subsequently converted into a baryon asymmetry by electroweak sphaleron processes. For a recent series of review articles on leptogenesis, see~\cite{Dev:2017trv,Drewes:2017zyw,Dev:2017wwc,Biondini:2017rpb,Chun:2017spz,Hagedorn:2017wjy}.


An intrinsic limitation of standard thermal leptogenesis is that it requires a very high RHN mass scale. In the simplest scenario, sometimes referred to as vanilla leptogenesis, one finds, e.g., an absolute lower bound on the mass of the lightest RHN of about $M_1^{\rm min} \simeq 10^9\,\textrm{GeV}$~\cite{Davidson:2002qv,Buchmuller:2002rq,Giudice:2003jh,Buchmuller:2004nz}. Flavor effects allow to lower this bound by an additional order of magnitude, $M_1^{\rm min} \simeq 10^8\,\textrm{GeV}$~\cite{Blanchet:2008pw}, but not much further. The high RHN mass scale in standard thermal leptogenesis is due to the fact that the $CP$ asymmetry in RHN decays is proportional to the product of active and sterile neutrino masses. The tiny SM neutrino masses therefore necessitate large RHN masses, a relation that was first pointed out by Davidson and Ibarra (DI)~\cite{Davidson:2002qv}. 


A high RHN mass scale is problematic, or at least undesirable, for several reasons. First of all, RHN masses far above the electroweak scale preclude the possibility of directly probing the dynamics of leptogenesis in future collider experiments~\cite{Chun:2017spz}. Second, in the type-I seesaw model, the RHNs contribute to the renormalization group running of the SM Higgs mass parameter $\mu^2$. For large RHN masses, a $\mu^2$ parameter around the electroweak scale is therefore necessarily fine-tuned, which may be regarded as a naturalness problem~\cite{Vissani:1997ys}. And third, a future detection of lepton number violation at low energies may readily rule out high-scale leptogenesis altogether~\cite{Deppisch:2013jxa,Deppisch:2015yqa,Harz:2015fwa}. Taken together, these observations serve as a motivation to seek alternatives to the paradigm of standard thermal leptogenesis in the type-I seesaw model that manage to generate the BAU at a (much) lower RHN mass scale.


The main purpose of this paper is to present a promising example for such a low-scale alternative to standard thermal leptogenesis. To this end, we will carry out an in-depth study of thermal leptogenesis in Ernest Ma's scotogenic model of radiative neutrino masses~\cite{Ma:2006km}. This model, which is arguably the simplest model of radiative neutrino masses, is particularly attractive as it unifies the generation of SM neutrino masses with the physics of dark matter (DM). The scotogenic model enlarges the SM field content by a second $SU(2)_L$ scalar doublet $\eta$ and at least two RHNs $N_i$, all of which are supposed to transform odd under an exact $\mathbb{Z}_2$ symmetry. The $\mathbb{Z}_2$ symmetry serves two purposes. It stabilizes the lightest $\mathbb{Z}_2$-odd state, such that it becomes a good DM candidate, if it is electrically neutral, and it prevents the $\eta$ doublet from obtaining a nonzero vacuum expectation value (VEV), so that no neutrino masses can be generated at tree level. This renders the SM neutrino masses \textit{scotogenic}, i.e., they only arise via radiative corrections in the dark sector. In this way, the scotogenic model offers a natural explanation for the suppressed masses of the active SM neutrinos.


Leptogenesis can proceed via a variety of mechanisms in the scotogenic model, depending on the details of the mass spectrum in the $\mathbb{Z}_2$-odd sector. Here, an important question is the choice of the mass eigenstate that is supposed to account for DM. In principle, one faces two options. DM can either be fermionic and consist of the lightest RHN $N_1$~\cite{Kubo:2006yx,Sierra:2008wj,Suematsu:2009ww} or it can be bosonic and consist of the lightest neutral component in the scalar $\eta$ doublet~\cite{LopezHonorez:2006gr,Hambye:2009pw,Dolle:2009fn,Honorez:2010re, Goudelis:2013uca, Krawczyk:2013jta, Diaz:2015pyv,Garcia-Cely:2015khw, Borah:2017dfn}. In the former case, the DM relic density is sensitive to the neutrino Yukawa couplings, while in the latter case, it mostly depends on the scalar and gauge interactions of the particles in the $\eta$ multiplet. As it turns out, DM in the form of RHNs typically requires large Yukawa couplings, which implies an efficient washout of lepton asymmetry during leptogenesis~\cite{Ma:2006fn}. In the fermionic DM scenario, it is therefore impossible to realize ordinary thermal leptogenesis via the decay of RHNs with a hierarchical mass spectrum. Instead, one has to resort to alternative mechanisms, such as, e.g., resonant leptogenesis~\cite{Pilaftsis:1997jf,Pilaftsis:2003gt} (see~\cite{Suematsu:2011va} for an explicit study). Similarly, it is possible to generate the BAU via the Akhmedov-Rubakov-Smirnov mechanism of RHN oscillations~\cite{Akhmedov:1998qx} and/or via $CP$-violating Higgs decays~\cite{Hambye:2016sby,Hambye:2017elz} in the fermionic DM case (see~\cite{Baumholzer:2018sfb} for a recent study). Together with the requirement to reproduce the DM relic abundance, all these alternative realizations of thermal leptogenesis require some degree of degeneracy in the RHN mass spectrum. This corresponds to an extra physical assumption which needs to be justified by an additional theoretical ingredient (such as, e.g., a flavor symmetry). However, \textit{a priori}, the scotogenic model does not require any such additional assumption to explain the low-energy neutrino data. Therefore, we shall ignore the possibility of resonant leptogenesis and focus on the case of hierarchical RHNs in the following. In addition, we recall that the large Yukawa couplings in the fermionic DM case easily lead to a violation of constraints on lepton flavor violation~\cite{Adulpravitchai:2009gi,Toma:2013zsa,Vicente:2014wga}. For these reasons, we will settle for the second option and assume that DM consists of scalar $\eta$ particles in this paper. In summary, this means that we will consider a mass spectrum of the form,
\begin{align}
\label{eq:spectrum}
M_{i+1} \gtrsim 3 \,M_i \,,\quad M_1 \gg m_\eta \,,
\end{align}
where $i=1$ or $i=1,2$ and where $m_\eta$ denotes the mass of the $\eta$ multiplet before electroweak symmetry breaking.


There exist various studies of thermal leptogenesis in the scotogenic model in the literature. Ma himself was the first to point out that the scotogenic model could serve as a simultaneous explanation of SM neutrino masses, DM, and the BAU~\cite{Ma:2006fn}. More detailed studies were subsequently presented in~\cite{Kashiwase:2012xd,Kashiwase:2013uy,Racker:2013lua,Clarke:2015hta}. However, it seems that none of these studies is fully exhaustive. The analyses in~\cite{Kashiwase:2012xd,Kashiwase:2013uy}, e.g., focus on very particular choices for the neutrino Yukawa couplings motivated by the experimental data on the neutrino mixing angles. However, they neglect all flavor effects in the computation of the lepton asymmetry. This does not really capture the essence of the problem, since unflavored leptogenesis is actually independent of the parameters in the lepton mixing matrix~\cite{Dev:2017trv} (see also Eq.~\eqref{eq_h_dagger_h} below). In any case, it is evident that the studies in~\cite{Kashiwase:2012xd,Kashiwase:2013uy} only cover a small part of the available parameter space. Meanwhile, the analysis in~\cite{Racker:2013lua} is mostly concerned with the study of general parameter relations in the RHN sector. It does not intend to reproduce the neutrino oscillation data and, thus, only incorporates an order-of-magnitude estimate of the active neutrino mass scale. But more importantly, it derives all numerical results in the limit of only two RHNs. This is surprising, as one can show on rather general grounds that the predictions of the scotogenic model in the two-RHN (2RHN) limit do not substantially differ from those in the type-I seesaw model (see our discussion in Sec.~\ref{sec_twoRHneutrinos}). Moreover, the three studies in~\cite{Kashiwase:2012xd,Kashiwase:2013uy,Racker:2013lua} all resort to a resonant enhancement of the $CP$ asymmetry at one point or another. As we will see in this paper, this is actually not necessary for RHN masses down to $M_1^{\rm min} \sim 10\,\textrm{TeV}$. Finally, the analysis in~\cite{Clarke:2015hta} studies thermal leptogenesis in two-Higgs-doublet models from a more general perspective. It properly accounts for the low-energy neutrino data, but only provides a few analytical estimates and refrains from actually solving the corresponding set of Boltzmann equations. A more comprehensive summary of leptogenesis studies in extensions of the scotogenic model as well as other scotogenic-like models of radiative neutrino masses can be found in~\cite{Cai:2017jrq}.


The above considerations motivate us to revisit thermal leptogenesis via the decay of hierarchical RHNs in the scotogenic model and to re-evaluate the important question as to what extent the leptogenesis scale can be lowered in this model. In doing so, we will attempt to present transparent analytical arguments wherever possible.


The remainder of this paper is organized as follows. In the next section, we will introduce the scotogenic model and summarize its key features. In Sec.~\ref{sec_ingredients}, we will then collect all expressions that are necessary to study leptogenesis in the scotogenic model. Next, in Sec.~\ref{sec_twoRHneutrinos}, we will first discuss the case of two RHNs. This will lead us to the important conclusion that the 2RHN case does not really allow for any improvement over standard thermal leptogenesis in the type-I seesaw model. In Sec.~\ref{sec_threeRHneutrinos}, we will finally turn to the core of our analysis and present a detailed analytical and numerical discussion of the three-RHN (3RHN) case. Sec.~\ref{sec_conclusion} contains our conclusions. 

\section{The Scotogenic Model}\label{sec_scotogenicModel}

We begin by summarizing the main properties of Ma's scotogenic model of radiative neutrino masses~\cite{Ma:2006km}. The new fields in this model are two or more RHNs $N_i$ as well as an inert Higgs doublet $\eta$. The interaction Lagrangian of these fields is reminiscent of the type-I seesaw scenario,
\begin{equation}
\label{eq_Lagrangian}
\mathcal{L}_{N,\eta} = -h_{\alpha i}\,\overline{\ell_L^\alpha} \, \tilde{\eta}\,N_i + \frac{1}{2} M_i \overline{N_i} (N^c)_i + \mathrm{h.c.},
\end{equation}
with the Yukawa couplings $h_{\alpha i}$, the SM lepton doublets $\ell_L^\alpha \equiv (\nu_L^\alpha, \alpha_L)^T$ ($\alpha = e,\mu,\tau$), the conjugate scalar doublet $\tilde{\eta} \equiv i \sigma_2 \eta^*$ and the Majorana masses $M_i$. Note that $\left<\eta\right> \neq 0$ would break the $\mathbb{Z}_2$ symmetry. Therefore, unlike in the seesaw scenario, no Dirac mass term is generated upon electroweak symmetry breaking. The scalar sector of the model includes the SM Higgs doublet $H$ as well as the inert doublet $\eta$ and is described by the potential
\begin{align}
V(H, \eta) &= \mu^2H^\dagger H + m_\eta^2 \eta^\dagger \eta + \frac{\lambda_1}{2} (H^\dagger H)^2 + \frac{\lambda_2}{2} (\eta^\dagger \eta)^2 \nonumber\\
&\quad + \lambda_3 (H^\dagger H) (\eta^\dagger \eta) + \lambda_4 (H^\dagger \eta) (\eta^\dagger H) \nonumber\\
&\quad + \frac{\lambda_5}{2} \left[ (H^\dagger \eta) (H^\dagger \eta) + (\eta^\dagger H) (\eta^\dagger H) \right],
\end{align}
where all $\lambda_i$ can be chosen real without loss of generality.


After electroweak symmetry breaking, the physical scalar states can be identified as $H = (0, (v+h)/\sqrt{2})^T$ and $\eta = (\eta^+, (\eta_R + i\eta_I)/\sqrt{2})^T$ with masses
\begin{equation}
\begin{aligned}
m_h^2 &= \lambda_1 v^2, \\
m_{\eta^\pm}^2 &= m_\eta^2 + \frac{v^2}{2} \lambda_3,\\
m_{\eta_R}^2 &= m_\eta^2 + \frac{v^2}{2} (\lambda_3 +\lambda_4 + \lambda_5),\\
m_{\eta_I}^2 &= m_\eta^2 + \frac{v^2}{2} (\lambda_3 +\lambda_4 - \lambda_5),
\end{aligned}
\end{equation}
where the SM Higgs doublet VEV $v=246 \,\text{GeV}$ appears. In the following, we shall assume that $\lambda_4 \pm \lambda_5 < 0$ and $\lambda_5 > 0$. In this case, both the real scalar $\eta_R$ and the real pseudoscalar $\eta_I$ are lighter than the complex scalar $\eta^\pm$. Moreover, given our assumptions in Eq.~\eqref{eq:spectrum}, $\eta_I$ turns out to be the lightest state in the entire dark matter sector. This renders $\eta_I$ the DM candidate in our model. The case of inert doublet DM in the scotogenic model is well studied, and it is found that, for the mass range~\cite{LopezHonorez:2006gr,Hambye:2009pw,Dolle:2009fn,Honorez:2010re, Goudelis:2013uca, Krawczyk:2013jta, Diaz:2015pyv,Garcia-Cely:2015khw, Borah:2017dfn}
\begin{align}
\label{eq:DMmassbound}
534\GeV \leq m_{\eta_I} \lesssim 20\TeV \,,
\end{align}
the correct relic abundance can be achieved, while all constraints are evaded by adjusting the scalar couplings accordingly. On the other hand, for the purposes of leptogenesis, the only relevant scalar coupling turns out to be $\lambda_5$. 


Given the fields and couplings introduced above, the active neutrino mass matrix turns out to be~\cite{Ma:2006km,Merle:2015ica}
\begin{align}
\label{eq_active_neutrino_masses}
\left(\mathcal{M}_\nu\right)_{\alpha \beta} = \sum_i\frac{M_i h_{\alpha i}^* h_{\beta i}^*}{32 \pi^2}
\left[L\left(m_{\eta_R}^2\right) - L\left(m_{\eta_I}^2\right) \right],
\end{align}
where the function $L$ helps us to simplify our notation,
\begin{align}
L\left(m^2\right) \coloneqq \frac{m^2}{m^2-M_i^2}\, \lnx{\frac{m^2}{M_i^2}} \,.
\end{align}
We note that $m_{\eta_R}^2 - m_{\eta_I}^2 = v^2 \lambda_5$ and, hence, the two real scalars $\eta_R$ and $\eta_I$ become degenerate in the limit $\lambda_5 \rightarrow 0$. The masses in Eq.~\eqref{eq_active_neutrino_masses} then vanish and one can define a global $U(1)$ lepton number symmetry. Therefore, $\lambda_5$ is a naturally small coupling in the sense of 't Hooft~\cite{tHooft:1979rat}.


It is convenient to introduce an adapted Casas-Ibarra (CI) parametrization~\cite{Casas:2001sr} for the Yukawa matrix $h$. For this purpose, we rewrite Eq.~\eqref{eq_active_neutrino_masses} in matrix form as
\begin{equation}
\mathcal{M}_\nu = h^* \Lambda^{-1} h^\dagger,
\end{equation}
where we introduced the diagonal matrix $\Lambda$ with entries
\begin{equation}
\label{eq_def_Lambda}
\Lambda_i \coloneqq \frac{2 \pi^2}{\lambda_5}\,\xi_i\,\frac{2 M_i}{v^2}
\end{equation}
and
\begin{align}
\xi_i \coloneqq \left(\frac{1}{8} \frac{M_i^2}{m_{\eta_R}^2 - m_{\eta_I}^2}
\left[ L\left(m_{\eta_R}^2\right) - L\left(m_{\eta_I}^2\right)\right] \right)^{-1}.
\end{align}
The parameters $\xi_i$ are of order one in most of the parameter space of interest. Note that we split the inverse mass scales $\Lambda_i$ into factors which are also present in the type-$\rom{1}$ seesaw, $2 M_i/v^2$, and additional factors which are characteristic of the scotogenic model, $(2 \pi^2/\lambda_5)\,\xi_i$. Following the notation of~\cite{Casas:2001sr}, the in general complex symmetric mass matrix $\mathcal{M}_\nu$ is diagonalized by the Pontecorvo-Maki-Nakagawa-Sakata (PMNS) leptonic mixing matrix $U$~\cite{Pontecorvo:1957qd,Maki:1962mu} via $D_\nu = U \mathcal{M}_\nu U^T$, and we find that the Yukawa couplings can be written as
\begin{align}
\label{eq_Yukawas_CI_parametrization}
h_{\alpha i} = \left( U\,D_{\sqrt{\mathcal{M}_\nu}}\, R^\dagger\, D_{\sqrt{\Lambda}} \right)_{\alpha i},
\end{align}
where the arbitrary complex matrix $R$ satisfies $RR^T = 1$.

\section{Ingredients for leptogenesis}\label{sec_ingredients}

We now turn to the discussion of thermal leptogenesis in the scotogenic model. Our main goal in this paper will be to gain an \textit{analytical} understanding of leptogenesis in the scotogenic model. In particular, we wish to highlight the relevant parameter relations that eventually result in the observed BAU. As we will see, this will provide us with new and valuable insights regarding the interplay of the active and sterile neutrino masses that significantly extend the existing results in the literature. Therefore, to keep the discussion clear and concise and to facilitate the analytical treatment, we will restrict our analysis to only the most important physical effects. That is, we will focus on the decays and inverse decays of $N_1$ neutrinos as well as on the corresponding $\Delta L =2$ washout processes. The asymmetries generated in $N_{2,3}$ decays together with any preexisting $B\!-\!L$ asymmetry are negligible because of strong washout effects either mediated by the $N_1$ or the $N_{2,3}$ themselves. Accordingly, the initial or previously generated asymmetry is almost entirely washed out and only the $N_1$ contribution survives. Possible corrections to our analysis (which we will neglect) include $\Delta L = 1$ scatterings~\cite{Luty:1992un,Plumacher:1996kc}, thermal corrections~\cite{Giudice:2003jh,Kiessig:2010pr}, flavor effects~\cite{Blanchet:2011xq,Antusch:2010ms} and quantum kinetic effects~\cite{Buchmuller:2000nd,DeSimone:2007gkc}. A more comprehensive analysis taking into account some or even all of these effects is left for future work.


Let us now collect the various expressions and quantities that are necessary to describe thermal leptogenesis in the scotogenic model. Our conventions and notation are based on~\cite{Buchmuller:2004nz,Antusch:2010ms}. The analytical relations from~\cite{Buchmuller:2004nz} that we will use in this paper are also valid for our model because the underlying Boltzmann equations turn out to be identical (cf. Sec.~\ref{subsec_numericalInsights}). As in standard thermal leptogenesis, we have to distinguish between a weak washout and a strong washout regime. The different regimes are characterized by different values of the decay parameter
\begin{equation}
\label{eq_def_decay_parameter}
K_1 \coloneqq \frac{\Gamma_1}{H(z_1=1)},
\end{equation}
with the $N_1$ decay width $\Gamma_1$, the Hubble parameter $H$ and $z_1 \coloneqq M_1/T$ with temperature $T$ of the photon bath. Leptogenesis occurs above the electroweak scale during the era of radiation domination. The Hubble parameter can therefore be expressed in terms of $T$ as follows,
\begin{equation}
\label{eq_Hubble_parameter}
H = \sqrt{\frac{8 \pi^3 g_*}{90}} \frac{T^2}{M_\text{Pl}} = H(z_1=1)\,\frac{1}{z_1^2},
\end{equation}
where $g_*$ is the effective number of relativistic degrees of freedom\footnote{The effective number of relativistic degrees of freedom $g_*$ is given by $g_* = 114.25$ for two RHNs and $g_* = 116$ for three RHNs.} and $M_\text{Pl} \simeq 1.22 \cdot 10^{19}\,\textrm{GeV}$ the Planck mass. The regimes that are typically distinguished are the weak washout regime for $K_1 \lesssim 1$ and the strong washout regime for $K_1 \gtrsim 4$, with a transition region in between.


Next, we calculate the $CP$ asymmetry parameter $\varepsilon$ for $N_i \rightarrow \ell_L^\alpha \eta,\overbar{\ell_L^\alpha} \eta^*$ decays, which leads us to
\begin{align}
\label{eq_CPasymmetry}
\varepsilon_{i \alpha} & = \frac{1}{8\pi (h^\dag h)_{ii}} \sum_{j\neq i} \Bigg[ f\!\left(\frac{M_j^2}{M_i^2},\frac{m_\eta^2}{M_i^2}\right) \Im{h_{\alpha i}^*h_{\alpha j} (h^\dag h)_{ij}}
\nonumber \\
&\quad -\frac{M_i^2}{M_j^2-M_i^2} \left(1-\frac{m_\eta^2}{M_i^2}\right)^2 \Im{h_{\alpha i}^*h_{\alpha j} H_{ij}} \Bigg].
\end{align}
In this expression, the function $f$ originates from the interference of the tree-level diagram with the one-loop vertex correction and is given by
\begin{align}
f(&r_{j i},\eta_i) \coloneqq
\nonumber \\
&\sqrt{r_{j i}} \bigg[ 1 + \frac{(1 - 2 \eta_i + r_{j i})}{(1 - \eta_i)^2}
\lnx{\frac{r_{j i} - \eta_i^2}{1 - 2 \eta_i + r_{j i}}}\bigg],
\end{align}
with $r_{j i} \coloneqq M_j^2 / M_i^2$ and $\eta_i \coloneqq m_\eta^2 / M_i^2$. In the limit of $m_\eta=0$ this reduces to the well-known result~\cite{Covi:1996wh}
\begin{equation}
  f(r_{j i},0) = \sqrt{r_{j i}} \left[1 + (1 + r_{j i}) \lnx{\frac{r_{j i}}{r_{j i} + 1}}\right].
\end{equation}
Similarly, we obtain for the self-energy contributions
\begin{equation}
H_{ij} \coloneqq (h^\dag h)_{ij} \frac{M_j}{M_i} + (h^\dag h)_{ij}^*.
\end{equation}
However, if we neglect flavor effects and therefore sum over the final state flavor $\alpha$, the second term in $H_{ij}$ may be omitted since it will not contribute to the imaginary part in Eq.~\eqref{eq_CPasymmetry}. In this case, we obtain the simpler expression
\begin{equation}
\label{eq_CPasymmetry_flavor_summed}
\varepsilon_{i} = \frac{1}{8\pi (h^\dag h)_{ii}} \sum_{j \neq i} \Im{(h^\dag h)_{ij}^2} \frac{1}{\sqrt{r_{j i}}} \, F(r_{j i}, \eta_i)
\end{equation}
where we defined
\begin{equation}
F(r_{j i}, \eta_i) \coloneqq \sqrt{r_{j i}}  \left[ f(r_{j i}, \eta_i) - \frac{\sqrt{r_{j i}}}{r_{j i} - 1} (1 - \eta_i)^2 \right].
\end{equation}
Furthermore, the decay width $\Gamma_1$ that appears in the decay parameter $K_1$ can be calculated to be
\begin{equation}
\label{eq_decay_width_tree_level}
\Gamma_1 = \frac{M_1}{8 \pi} \left( h^\dagger h \right)_{1 1} (1 - \eta_1)^2.
\end{equation}


Finally, before we turn to the different cases of two and three RHNs, it is worth having a closer look at the frequently appearing expression $h^\dagger h$. Using the CI parametrization, we find from Eq.~\eqref{eq_Yukawas_CI_parametrization} that
\begin{equation}
\label{eq_h_dagger_h}
\left( h^\dagger h \right)_{i j} = \sqrt{\Lambda_i \Lambda_j} \left( R D_{\mathcal{M}_\nu} R^\dagger \right)_{i j}.
\end{equation}
Here, $\widetilde{m} \coloneqq R D_{\mathcal{M}_\nu} R^\dagger$ only depends on the masses of the active neutrinos through $D_{\mathcal{M}_\nu} \coloneqq \diag{m_1,m_2,m_3}$ and the (complex) CI parameters, whereas the dependence on other parameters like $M_i$ and $\lambda_5$ appears by means of $\Lambda_i$. Interestingly enough, the matrix $h^\dagger h$ is independent of the PMNS matrix $U$. This indicates that the $CP$-violating phases relevant for leptogenesis are independent of the $CP$-violating phases in the PMNS matrix.%
\footnote{This situation changes once flavor effects are taken into account. In this case, the $CP$ asymmetry parameters $\varepsilon_{i\alpha}$ also depend on factors of the form $h_{\alpha i}^*h_{\alpha j}^{\vphantom{*}}$ (see Eq.~\eqref{eq_CPasymmetry}) that are sensitive to the $CP$-violating phases
in the lepton sector at low energies.}
Similarly, it shows that unflavored leptogenesis is insensitive to the values of the neutrino mixing angles (see our discussion in the introduction regarding the analyses in~\cite{Kashiwase:2012xd,Kashiwase:2013uy}).

\section{Two Right-Handed Neutrinos}\label{sec_twoRHneutrinos}

Using the formulas from the previous section and specifying them to the case of two RHNs, we can check whether low-scale leptogenesis is feasible in this scenario. With two RHNs, only two active neutrinos obtain a nonzero mass and we distinguish between normal ordering (NO) and inverted ordering (IO), 
\begin{align}
D_{\mathcal{M}_\nu}^\text{NO} & =
\diag{0, \sqrt{\Delta m_{2 1}^2}, \sqrt{\Delta m_{3 1}^2}} 
\,, \nonumber \\
D_{\mathcal{M}_\nu}^\text{IO} & =
\diag{\sqrt{-\Delta m_{3 1}^2}, \sqrt{\Delta m_{2 1}^2 - \Delta m_{3 1}^2}, 0} \,.
\end{align}
where $\Delta m_{i j}^2 \coloneqq m_i^2 - m_j^2$. To avoid duplicating equations, we introduce the notation $m_h$ for the heaviest active neutrino and $m_l$ for the lightest (massive) active neutrino.


In the case of only two RHNs, the matrix $R$ in Eq.~\eqref{eq_Yukawas_CI_parametrization} becomes a function of only one complex rotation parameter $z = z_R + i z_I$~\cite{Ibarra:2003up}, where $z_R \in \left[0,2\pi\right)$ and $z_I \in \mathbb{R}$. We can therefore readily maximize the $CP$ asymmetry in Eq.~\eqref{eq_CPasymmetry_flavor_summed} over all possible values of $z$,
\begin{equation}
|\varepsilon_1| \lesssim \frac{3 \pi}{4 \lambda_5 v^2}\,\xi_2 \left( m_h - m_l \right) M_1 .
\end{equation}
Here, we used that $|F(r_{2 1}, \eta_1)| \lesssim 3/2$ for a hierarchical RHN mass spectrum. This is essentially the DI bound~\cite{Davidson:2002qv}, except for the additional factor $(2 \pi^2 / \lambda_5) \,\xi_2$ (see Eq.~\eqref{eq_def_Lambda}). It is interesting to note that the factor $m_h - m_l$ suggests that, in the 2RHN case, the $CP$ asymmetry parameter can be larger for NO than for IO.


The final baryon-to-photon ratio $\eta_B = -C \varepsilon_1 \kappa_1$ follows from $\varepsilon_1$ after multiplication with an efficiency factor $\kappa_1$, which accounts for the effect of washout, and a conversion factor $C \simeq 0.01$~\cite{Buchmuller:2004nz}, which accounts for sphaleron conversion and entropy production after the generation of the lepton asymmetry. The efficiency factor $\kappa_1$ is a function of the decay parameter $K_1$, for which we obtain
\begin{align}
\label{eq_decay_parameter_1}
K_1 = \frac{2 \pi^2}{\lambda_5} \xi_1 \sqrt{\frac{45}{64 \pi^5 g_*}} \frac{M_{\rm Pl}}{v^2}  \widetilde{m}_{1 1} (1 - \eta_1)^2.
\end{align}
For $0 \leq \lambda_5 \leq 4 \pi$ and $3\,m_\eta \leq M_1$ (or $\eta_1 \leq 1/9$), we find that $K_1$ cannot become smaller than $K_1^{\rm min} \simeq 10$. For large parts of the parameter space, $K_1$ is even significantly larger, $ K_1\sim 10^3$ and above. We are therefore always in the strong washout regime. This means that we can safely assume $N_1$-dominated leptogenesis and neglect washout through scattering effects. Additionally, the large value of $K_1$ also enables us to use the approximation for the efficiency factor in the strong washout regime~\cite{Buchmuller:2004nz},
\begin{equation}
\label{eq_strong_washout_efficiency_factor}
\kappa_1(K_1) \simeq \frac{1}{1.2 K_1 \left[\ln K_1\right]^{0.8}}.
\end{equation}
Taking everything together, we find
\begin{align}
\begin{aligned}
& \eta_B = C \sqrt{\frac{16 \pi^3 g_*}{1.2^2 \cdot 45}} \frac{M_1}{M_{\rm Pl}} \frac{\xi_2}{\xi_1} \frac{F(r_{2 1}, \eta_1)}{(1 - \eta_1)^2} \frac{1}{\left[\ln K_1\right]^{0.8}} \cdot \\
& \frac{\left(m_h^2 - m_l^2\right) \sin(2 z_R) \sinh(2 z_I)}{[-(m_h - m_l) \cos(2 z_R) + (m_h + m_l) \cosh(2 z_I)]^2},
\end{aligned}
\end{align}
where $K_1$ inside the logarithm is again given by the expression in Eq.~\eqref{eq_decay_parameter_1}. One interesting aspect of this formula is that $\eta_B$ only depends logarithmically on $\lambda_5$ (through $K_1$), which means that it has only a minor influence on the generated baryon asymmetry. Furthermore, since $K_1 \sim 1 / \lambda_5$, the baryon asymmetry $\eta_B$ actually decreases if $\lambda_5$ is decreased; so the perturbative limit $\lambda_5 = 4 \pi$ allows for the largest generated baryon asymmetry. Remarkably enough, this is contrary to the naive expectation that a smaller value of $\lambda_5$ would result in a larger baryon asymmetry in consequence of a larger $CP$ asymmetry. Let us now fix the scalar coupling $\lambda_5$ at $4\pi$ and maximize $\eta_B$ over the CI parameters $z_R$ and $z_I$. Using the experimentally measured value $\eta_B^{\rm obs} \simeq 6.1 \cdot 10^{-10}$, we can then derive a lower limit on the mass $M_1$ of lightest RHN that still allows for successful leptogenesis,
\begin{equation}
M_{1, \,\text{min}}^\text{NO} \sim 10^{10} \GeV \quad \text{and} \quad M_{1, \,\text{min}}^\text{IO} \sim 10^{12} \GeV.
\end{equation}
These numbers are basically identical to those that one obtains in the strong washout regime ($K_1 \gtrsim 10$) of standard type-$\rom{1}$ leptogenesis. The reason that in this scenario not much changes is that, in the strong washout regime, we have $\eta_B \sim \varepsilon_1 / K_1 \sim \Im{(h^\dag h)^2} / (h^\dag h)^2$ (neglecting logarithmic dependencies) and thus all the prefactors that enter into $h^\dagger h$, like $\lambda_5$, cancel. We conjecture that this is a generic feature of radiative neutrino mass models in which the active neutrino mass matrix is only modified by more or less simple multiplicative factors. However, this statement relies on being in the strong washout regime, which can depend, along other factors, on the mass of the lightest active neutrino, as we will see in the next part, when looking at the 3RHN case.

\section{Three Right-Handed Neutrinos}\label{sec_threeRHneutrinos}

The most crucial difference between the 3RHN case and the 2RHN case is that, with three RHNs, we are not necessarily in the strong washout regime. This is due to the fact that, in this case, the CI parametrization $R$ has three instead of one free parameter\,---\,it can be written as the product of three complex rotation matrices $R(z_{2 3})$, $R(z_{1 3})$, and $R(z_{1 2})$\,---\,and thus the accessible parameter space is vastly extended. Furthermore, we make two important observation: First, the factor of $(2 \pi^2 / \lambda_5) \,\xi_1$, which is peculiar to the scotogenic model, can only increase $K_1$ in the relevant parameter space (see~Eq.~\eqref{eq_decay_parameter_1}). And second, an explicit calculation shows that $\widetilde{m}_{1 1} = \left( R D_{\mathcal{M}_\nu} R^\dagger \right)_{1 1} \geq m_l$. These insights, together with the values for $M_{\rm Pl}$, $v$, etc., imply that
\begin{equation}
K_1 \gtrsim 10^{3} \left( \frac{m_l}{\text{eV}} \right) (1 - \eta_1)^2.
\end{equation}
Consequently, we are in the strong washout regime as long as $m_l \gtrsim 10^{-3} \,\text{eV}$ and $\eta_1 \not\approx 1$ (or $m_\eta \not\approx M_1$). In this regime, we end up with the same result as in the 2RHN case for NO or standard  type-\rom{1} leptogenesis, $M_1^\text{min} \sim 10^{10} \GeV$. However, with three RHNs, the distinction between NO and IO disappears because the mass difference of the heaviest and lightest (massive) active neutrinos is (nearly) identical in both cases.


For the maximally possible $CP$ asymmetry, we find by explicit calculation that it is independent of $z_{2 3}$. Moreover, an explicit parameter scan indicates that the optimal values for NO are $z_{1 2} = 0$ and $z_{{1 3}_R} = \pm z_{{1 3}_I}$, where the relative sign of $z_{1 3_I}$ determines the overall sign of the $CP$ asymmetry. For IO, one similarly obtains $z_{1 2_R} = \pi / 2$, $z_{1 2_I} = 0$ and $z_{1 3_R} = \pi / 2 \pm z_{1 3_I}$. In close resemblance to the 2RHN case, we arrive at a DI-type bound,
\begin{equation}
\label{eq_maxCPasym3RHN}
|\varepsilon_1| \lesssim \frac{3 \pi}{4 \lambda_5 v^2}\,\xi_3 \left( m_h - m_l \right) M_1,
\end{equation}
which again includes a factor of $(2 \pi^2 / \lambda_5) \,\xi_3$ that derives from the structure of the active neutrino mass matrix in the scotogenic model.


To study leptogenesis in the 3RHN case, we split the task into an analytical and a numerical part. First, we will analytically solve the scenario without $\Delta L = 2$ washout, then determine when $\Delta L = 2$ washout starts to become important, and finally confirm and extend our analytical understanding with a numerical analysis.

\subsection{Analytical Insights}\label{subsec_analyticalInsights}

In the case with three RHNs and negligible $\Delta L = 2$ washout, we have to find a new solution strategy because we are no longer automatically in the strong washout regime for all possible values of $\lambda_5$ and therefore Eq.~\eqref{eq_strong_washout_efficiency_factor} does not hold anymore. However, we can always make use of the coupling $\lambda_5$ to achieve the maximally possible $\eta_B = -C \varepsilon_1 \kappa_1$. To do so, let us consider the connection between $\varepsilon_1$ and $K_1$ through $\lambda_5$. From Eqs.~\eqref{eq_decay_parameter_1} and~\eqref{eq_maxCPasym3RHN}, we find $K_1 \sim 1 / \lambda_5$ and $\varepsilon_1 \sim 1 / \lambda_5$ (which also holds in the non-optimized case), so $\varepsilon_1 \sim K_1$ and therefore $\eta_B \sim \kappa_1 (K_1) \, K_1$. This means that, in order to maximize the baryon asymmetry $\eta_B$, we have to choose $\lambda_5$ such that $\kappa_1 (K_1) \, K_1$ becomes maximal. Taking into account $\kappa_1 (K_1)$ as determined in~\cite{Buchmuller:2004nz}, we find for thermal as well as for vanishing initial $N_1$ abundance that, to a good approximation, $K_{1,\,\text{opt}} \approx 3$ maximizes $\eta_B$, corresponding to $\kappa_{1,\,\text{opt}} \approx 0.15$. This leaves only $\varepsilon_1$ to be optimized.


Recasting Eq.~\eqref{eq_decay_parameter_1} as $\lambda_{5,\,\text{opt}} (K_{1,\,\text{opt}})$ and using it in the expression for $\varepsilon_1$, we can optimize over the CI parameters. We find the same optimal CI parameters as for the DI-type bound (cf.~the discussion above Eq.~\eqref{eq_maxCPasym3RHN}), with $z_{1 3_I}$ now being fixed to the optimal value $z_{1 3_I} \approx \sqrt{m_l/\left(2 m_h\right)}$.\footnote{This is the largest allowed value of ${z_{1 3}}_I$ ensuring that the mass parameter $\widetilde{m}_{11}$ remains of $\mathcal{O}(m_l)$. This guarantees that $K_1$ is small while $\varepsilon_1$ is large. The same procedure applies to ${z_{12}}_I$; however, the $z_{12}$-dependent contribution to the $CP$-asymmetry $\varepsilon_1$ is sub-dominant compared to the $z_{13}$-dependent contribution. Thus, maximizing $\varepsilon_1$ while minimizing $\widetilde{m}_{11}$ yields $z_{12}=0$.\label{foot:angles}} Therefore, the optimal value of $\lambda_5$ after optimization over the CI parameters becomes 
\begin{align}
\label{eq_l5opt}
\lambda_{5,\,\text{opt}} \approx 4 \pi\,\xi_1 \left( \frac{m_l}{10^{-3}\,\text{eV}} \right) (1 - \eta_1)^2.
\end{align}
Neglecting $\eta_1$ and using $\xi_1 \sim 1$, we find that, for a lightest active neutrino mass $m_l > 10^{-3}\,\text{eV}$, i.e., in the strong washout regime, the optimal value for $\lambda_5$ is $4 \pi$, as expected from the 2RHN case. Larger $\lambda_5$ values would violate perturbativity and are, hence, not allowed.


On the other hand, the situation drastically changes if we consider one small active neutrino mass, $m_l^2 \ll m_h^2$, which allows us to explore the weak washout regime. In this case, we find for the upper limit of the baryon-to-photon ratio
\begin{equation}
\label{eq_etaB_3RHN}
\eta_B \lesssim 3 \cdot 10^{-21} \left(\frac{M_1}{\text{GeV}}\right) \frac{\xi_3}{\xi_1} \frac{m_h}{m_l}
\end{equation}
and therefore for the lower limit on the mass $M_1$,
\begin{equation}
\label{eq_minM1_3RHN}
M_1^\text{min} \approx \frac{\xi_1}{\xi_3} \frac{m_l}{m_h} \, 2 \cdot 10^{11} \GeV.
\end{equation}
This expression, which is valid for $m_l \lesssim 10^{-3} \,\text{eV}$ down to the point where $\Delta L = 2$ washout starts to become important, clearly shows the relevance of the mass of the lightest active neutrino. The reason why we can achieve lower values of $M_1^\text{min}$ compared to standard thermal leptogenesis is that, in the scotogenic model, the new parameter $\lambda_5$ enables us to go to the sweet spot of the $N_1$ decay parameter, $K_1 \rightarrow K_{1,\,\text{opt}} \approx 3$, while keeping all other parameters fixed. However, this optimization can only enlarge $K_1$ and cannot be continued to arbitrarily small values of $\lambda_5$. As we will see, using $\lambda_5$ in this way also increases the $\Delta L = 2$ washout~\cite{Clarke:2015hta}, which will, in combination with the electroweak sphalerons dropping out of equilibrium, eventually provide us with an absolute lower bound on $M_1$.


Next, let us discuss the effect of $\Delta L=2$ washout processes in the thermal bath. These processes consist of two-to-two scatterings, $\ell\eta \leftrightarrow \bar{\ell}\eta^*$ and $\ell\ell\leftrightarrow\eta^*\eta^*$, that are mediated by RHNs in the intermediate state and that violate lepton number by two units. To determine their effect on the final asymmetry, we first compute the rate of $\Delta L=2$ washout processes, $\Gamma_{\Delta L = 2}$, in units of $Hz$. To do so, we use the averaged matrix element squared from~\cite{Buchmuller:2004nz} and take into account the modified CI parametrization for the  Yukawa couplings in the scotogenic model in Eq.~\eqref{eq_Yukawas_CI_parametrization}. This procedure results in%
\footnote{A similar expression can be found in~\cite{Buchmuller:2004nz} and~\cite{Clarke:2015hta}. The authors of~\cite{Buchmuller:2004nz} use $v = 174 \GeV$, whereas we work with $v = 246 \GeV$.}
\begin{equation}
\label{eq_L2washout}
\Delta W = \frac{\Gamma_{\Delta L = 2}}{H\,z_1} = \frac{36 \sqrt{5} M_{Pl}}{\pi^{\frac{1}{2}} g_\ell \sqrt{g_*} v^4} \frac{1}{z_1^2} \frac{1}{\lambda_5^2} M_1 \overbar{m}_\xi^2.
\end{equation}
Here, we also assumed $\eta_1 \approx 0$, for simplicity. The factor $g_\ell = 2$ in Eq.~\eqref{eq_L2washout} counts the internal degrees of freedom per active neutrino $\nu_L^\alpha$ or per charged lepton $\alpha_L$. The effective mass parameter $\overbar{m}_\xi$ is defined as follows,
\begin{equation}
\begin {aligned}
\overbar{m}_\xi^2 &\coloneqq \sum_{i,\,j} \xi_i \xi_j \Re{(R D_{\mathcal{M}_\nu} R^\dagger)_{i j}^2} \\ &\approx 4 \xi_1^2 m_l^2 + \xi_2^2 m_{h_2}^2 + \xi_3^2 m_h^2,
\end{aligned}
\end{equation}
where the last line uses the previously determined optimal CI parameters. This expression for $\Delta W$ is identical to the result in~\cite{Buchmuller:2004nz}, only with an additional factor of $(2 \pi^2 / \lambda_5)^2$ for the scotogenic model in front and a slightly modified effective mass $\overbar{m}_\xi$.


Assuming that the $\Delta L = 2$ washout becomes important after the baryon asymmetry generation is finished, we can split the two different washout contributions and use
\begin{equation}
\label{eq_total_efficiency_factor}
\kappa_1^\text{tot} = \kappa_1\, e^{-\int_{z_B}^\infty \mathrm{d}z \,\Delta W}
\end{equation}
for the total efficiency factor. The time $z_B$ when the baryon asymmetry generation is finished has been calculated in~\cite{Buchmuller:2004nz} and is given by $z_B(K_1 \approx 3) \approx 3.5$. Since the case of vanishing initial $N_1$ abundance is similar but worse (lower $\kappa_1$) compared to the case of thermal initial abundance, we will focus on the latter from now on. We, however, caution that, for small $N_1$ couplings, a thermal initial abundance can no longer be generated via the usual inverse decays in the thermal bath. In this case, we have to assume some additional interactions at high temperatures that first yield a thermal $N_1$ abundance and then freeze out before the onset of leptogenesis. For our purposes, it will not be necessary to specify the exact nature of these additional interactions. As an example, we merely mention that an extra gauge interaction mediated by a heavy $Z'$ vector boson could possibly help establish a thermal $N_1$ abundance (see~\cite{Racker:2008hp} and references therein). Apart from that, we will simply use the assumption of thermal initial conditions as a working hypothesis.


Explicitly calculating the integral in Eq.~\eqref{eq_total_efficiency_factor} and taking into account Eq.~\eqref{eq_l5opt} with $\eta_1 \approx 0$ and Eq.~\eqref{eq_minM1_3RHN} for the smallest possible $M_1$, we obtain the following expression,
\begin{equation}
\int_{z_B}^\infty \mathrm{d}z \,\Delta W \approx \frac{9 \sqrt{5} M_{Pl} \cdot 10^{5} \GeV}{7 \pi^{\frac{5}{2}} g_l \sqrt{g_*} v^4} \frac{\text{eV}^2}{m_l m_h} \frac{1}{\xi_1 \xi_3} \,\overbar{m}_\xi^2.
\end{equation}
We expect that $\Delta L = 2$ washout becomes important as soon as $\int_{z_B}^\infty \mathrm{d}z \,\Delta W > 0.1$, which corresponds to a $10 \%$ decrease of $\kappa_1^\text{tot}$. If we further assume that $\xi_i \sim 1$, we find that $\Delta L = 2$ washout becomes relevant for
\begin{equation}
\label{eq_ml_approx_validity}
m_l \lesssim 10^{-6} \,\text{eV}.
\end{equation}


This critical $m_l$ value at which $\Delta L = 2$ washout becomes important is derived using a rigid relation between $\lambda_5$ and $m_l$ (see~Eq.~\eqref{eq_l5opt}). The numerical analysis in the next section will, however, show that the true critical value of $m_l$ is actually smaller by roughly one order of magnitude, $m_l \lesssim 10^{-7} \,\text{eV}$, while still approximately fulfilling Eq.~\eqref{eq_minM1_3RHN} for $M_1^{\text{min}}$. As we will see, this is mostly the outcome of a less rigid relation between $\lambda_5$ and $m_l$.

\subsection{Numerical Insights}\label{subsec_numericalInsights}

Complementary to our analytical calculations, we also perform a fully numerical analysis of leptogenesis in the scotogenic model. This will allow us to validate our analytical results and provide us with further insights. Let us consider the Boltzmann equations for $N_1$-dominated leptogenesis including the effect of $\Delta L = 2$ washout,
\begin{align}
\label{eq_Boltz1}
\frac{\mathrm{d} N_{N_1}}{\mathrm{d} z_1} &= -D_1 (N_{N_1} - N_{N_1}^{\text{eq}}) \\
\label{eq_Boltz2}
\frac{\mathrm{d} N_{B-L}}{\mathrm{d} z_1} &= -\epsilon_1 D_1 (N_{N_1} - N_{N_1}^\text{eq}) - W^\text{tot} N_{B-L},
\end{align}
with the $z_1$-dependent quantities
\begin{align}
D_1 &= K_1\,z_1\,\frac{\mathcal{K}_1(z_1)}{\mathcal{K}_2(z_1)} ,\\
W^\text{tot} &= W_1 + \Delta W ,\\
W_1 &= \frac{1}{4}\,K_1\,z_1^3\,\mathcal{K}_1(z_1) ,\\
N_{N_1}^\text{eq} &= \frac{z_1^2}{2}\,\mathcal{K}_2(z_1).
\end{align}
Here, $\mathcal{K}_i(z_1)$ denote the modified Bessel functions of the second kind and $\Delta W$ is given by Eq.~\eqref{eq_L2washout}. The final $B\!-\!L$ asymmetry $N_{B-L}^f$ can be converted to the baryon-to-photon ratio $\eta_B = C N_{B-L}^f$, with $C \approx 0.0088$.\footnote{The conversion factor is given by $C = 3/4 \, C_\text{sph} \, g_*^0/g_*$ with $C_\text{sph} = 8/23$, $g_*^0 = 43/11$ and $g_* = 116$. Note that, in the sphaleron conversion factor $C_\text{sph}$, we account for the presence of two Higgs doublets in our model. Likewise, the effective number of relativistic degrees of freedom $g_*$ accounts for the presence of three RHNs in our model at high temperatures.} One might worry that the asymmetry generated in the $\eta$--$\bar{\eta}$--sector through $N_1$ decays (cf.~\cite{Racker:2014yfa}) is not accounted for in the Boltzmann equations. However, for values of $\lambda_5 \gtrsim 10^{-4}$, as are relevant in our case, the $\eta \eta \leftrightarrow H H$ interactions mediated by the $\lambda_5$ coupling are strong enough for the asymmetry carried by the inert doublets to be negligible.

\begin{figure}
\begin{center}
\includegraphics[width=\columnwidth]{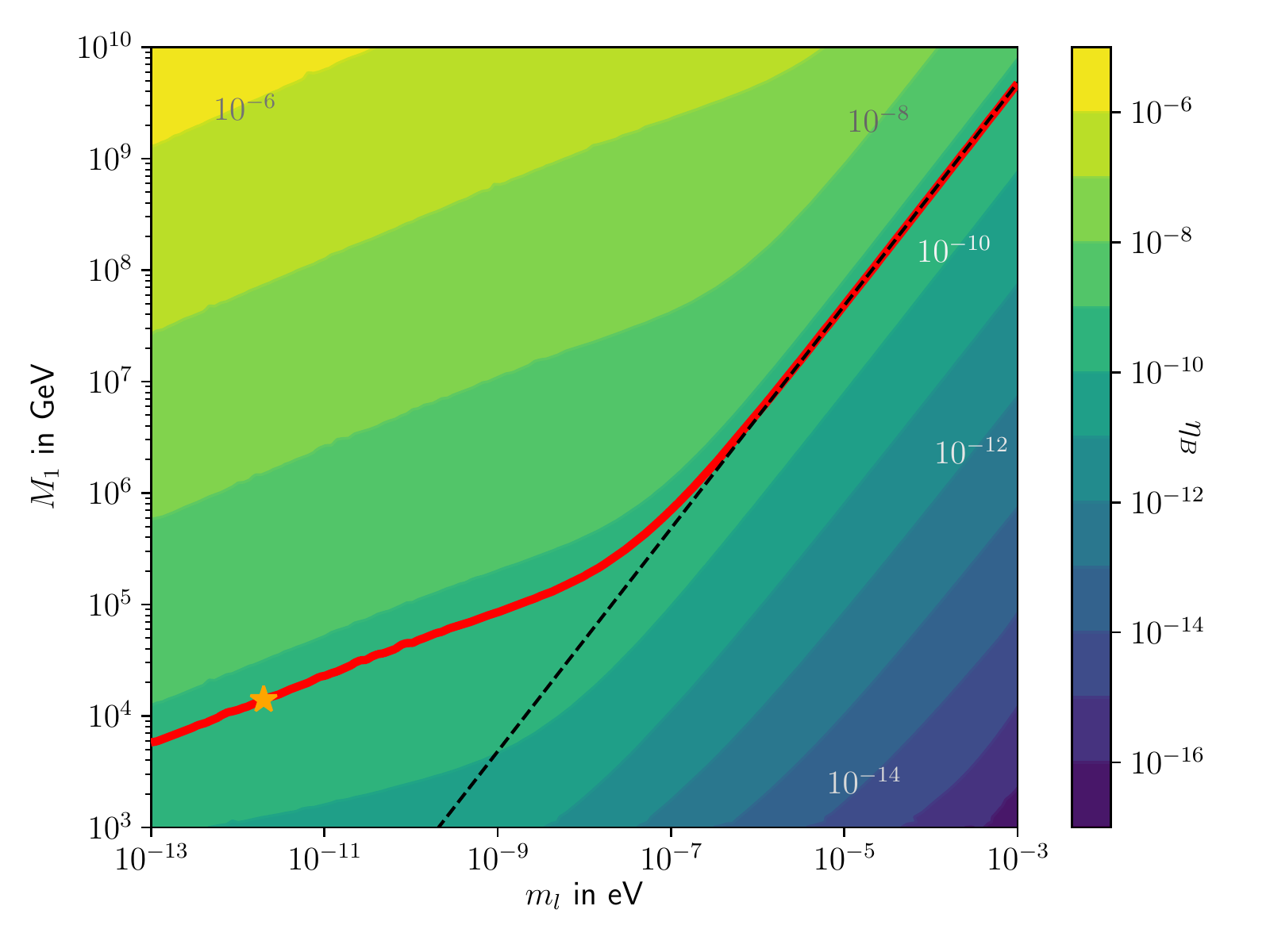}
\caption{Resulting baryon-to-photon ratio $\eta_B$ maximized over $\lambda_5 \in [10^{-6}, 4 \pi]$ of a parameter scan in the $m_l$--$M_1$ plane for $m_\eta = 550 \GeV$, $\lambda_3 = 1$, $\lambda_4 = -1$, $M_2 = 10^{0.5} M_1$, $M_3 = 10^{1} M_1$, $z_{1 2} = 0$, $z_{1 3_R} = \sqrt{m_l/\left(2 m_h\right)} = z_{1 3_I}$, $z_{2 3} = 0$ and NO. The red line shows the part of the scan that reproduces the observed value $\eta_B = 6.1 \cdot 10^{-10}$, the black line depicts the analytical solution given by Eq.~\eqref{eq_minM1_3RHN} with $\xi_1/\xi_3 \approx 1.2$ valid for $m_l \gtrsim 10^{-6} \,\text{eV}$ (cf. Eq.~\eqref{eq_ml_approx_validity}) and the orange star marks the lowest possible $M_1$ and $m_l$ for which the baryon asymmetry is generated before the $SU(2)$ sphalerons fall out of equilibrium.}
\label{fig_scan_results}
\end{center}
\end{figure}

Based on the Boltzmann equations~\eqref{eq_Boltz1} and~\eqref{eq_Boltz2}, we perform a parameter scan of leptogenesis in the scotogenic model, the result of which is shown in Fig.~\ref{fig_scan_results}. In this figure, we depict the maximally possible baryon-to-photon ratio $\eta_B$ in the $m_l$--$M_1$ parameter plane, since we are interested in the lowest possible $M_1$ and want to illustrate the strong dependency of $M_1^\text{min}$ on the active neutrino mass $m_l$. For all points in the plot with $\eta_B > \eta_B^\text{obs}$ (above the red line), the observed baryon-to-photon ratio can be achieved by appropriate choices of $\lambda_5$ and the CI parameters.


Using the numerical results from Fig.~\ref{fig_scan_results}, we can find an approximate expression for the upper limit of the baryon-to-photon ratio in the region where $M_1 \gg m_\eta$ and $\Delta L = 2$ scatterings are important, which can be compared to Eq.~\eqref{eq_etaB_3RHN} for the region where those scatterings are negligible. Approximately, we find
\begin{equation}
\eta_B \lesssim 1.6 \cdot 10^{-14} \:\bigg(\frac{m_l}{\text{eV}}\bigg)^{-0.19} \left(\frac{M_1}{\text{GeV}}\right)^{0.58}.
\end{equation}
The comparison with Eq.~\eqref{eq_etaB_3RHN}, which has exponents of $-1$ for $m_l$ and $1$ for $M_1$, shows a weaker dependence on both variables since the exponents are closer to zero. This is also reflected in a bigger spacing between the contours in the corresponding region of Fig.~\ref{fig_scan_results}.


\begin{figure}
\begin{center}
\includegraphics[width=\columnwidth]{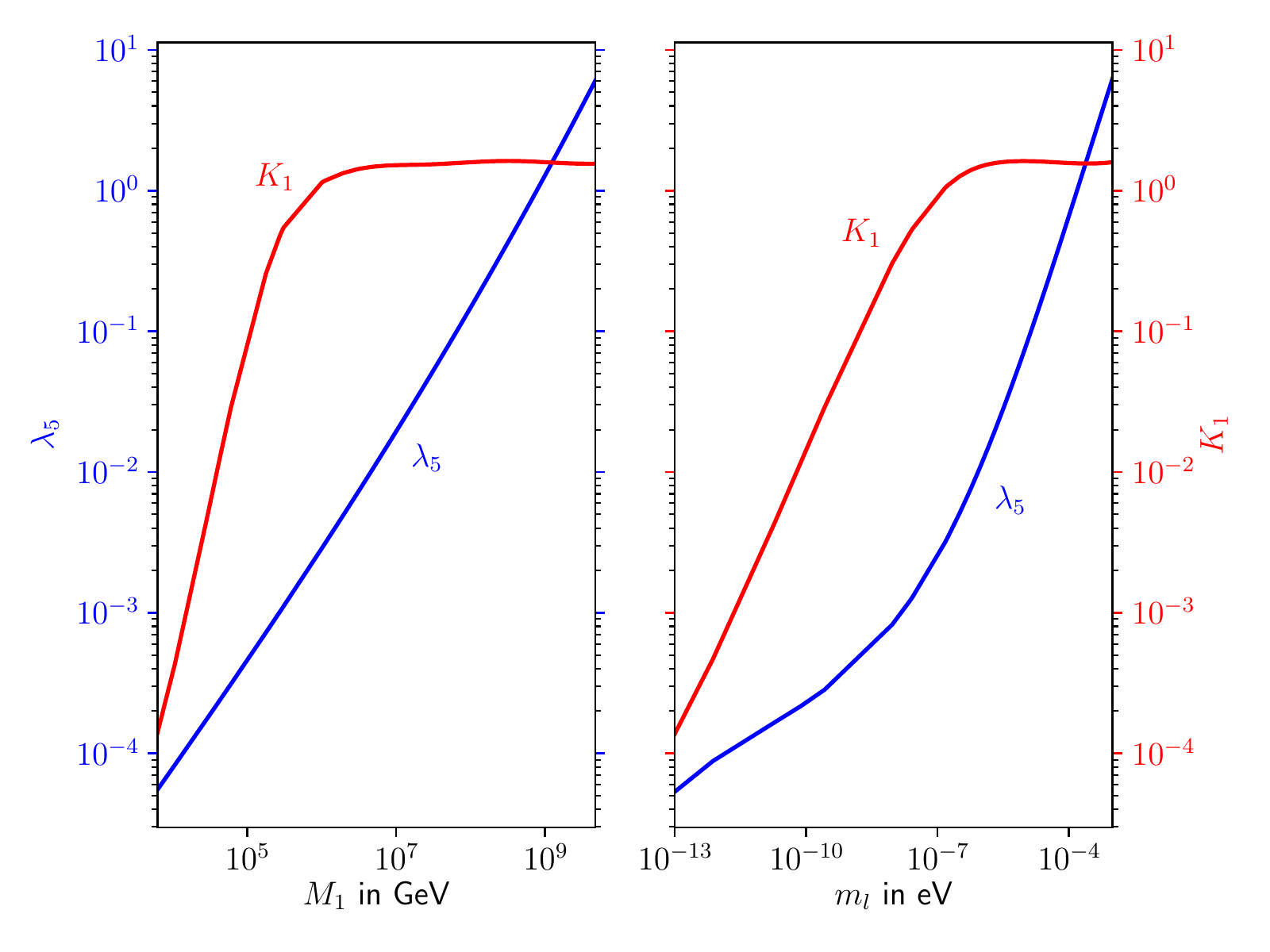}
\caption{Dependence of the scalar coupling $\lambda_5$ and the decay parameter $K_1$ on $M_1$ and $m_l$, respectively, for the observed baryon-to-photon ratio $\eta_B$ (cf. the red line in Fig.~\ref{fig_scan_results}).}
\label{fig_l5_K1_dependence}
\end{center}
\end{figure}

For further insights, Fig.~\ref{fig_l5_K1_dependence} gives an impression of how $\lambda_5$ and the decay parameter $K_1$ evolve in dependence of $M_1$ and $m_l$ for the correct baryon-to-photon ratio $\eta_B$ (cf. the red line in Fig.~\ref{fig_scan_results}). The transition between the regime in which $\Delta L = 2$ scatterings are negligible and where they are important is well visible as a kink in the decay parameter $K_1$. As long as $\Delta L = 2$ scatterings are negligible, $K_1$ is constant and of $\mathcal{O}(1)$, while we have to resort to the weak washout regime otherwise. The reason that the constant value of $K_1 \approx 1.5$ differs from $K_{1, \text{opt}} = 3$ as used in the analytical calculation, is that $K_{1, \text{opt}} = 3$ leads to a reasonably accurate expression for $\eta_B$ while enabling us to treat the two different cases of thermal and vanishing initial $N_1$ abundance simultaneously, which would not be possible for smaller values of $K_1$. Furthermore, Fig.~\ref{fig_l5_K1_dependence} also shows how the dependence of $\lambda_5$ on $m_l$ changes when $\Delta L=2$ scatterings become important around $m_l \approx 10^{-6} \,\text{eV}$, which is the ``less rigid relation'' mentioned at the end of Sec.~\ref{subsec_analyticalInsights}.


In the context of the small $\lambda_5$ values that appear in our scenario, one has to check that they are compatible with the constraints coming from direct detection via inelastic scattering. From Fig.~\ref{fig_l5_K1_dependence}, we can see that, for the observed baryon-to-photon ratio, we have $\lambda_5 \gtrsim 5 \cdot 10^{-5}$. Using the expression for the limit from direct detection via inelastic scattering from~\cite{Kashiwase:2013uy}, together with the experimental data from \textsc{XENON}100~\cite{Aprile:2011ts}, we find that, in our case, $\lambda_5$ has to be bigger than $3 \cdot 10^{-6}$, which is fulfilled.


In addition to the parameter scan with some fixed parameters, we also choose specific points in the $m_l$--$M_1$ plane that reproduce the measured $\eta_B$ and perform a complete scan over all CI parameters and $\lambda_5$. Since the complete scans do not show any possible significant improvement through the variation of other parameters, we are confident that Fig.~\ref{fig_scan_results} does, indeed, depict the lowest possible $M_1^\text{min}$. Small improvements are possible in the region where $\Delta L = 2$ washout is important. However, these improvements are not bigger than the systematic uncertainties of our simplified approach in which we neglect flavor effects, etc. (see the discussion in Sec.~\ref{sec_scotogenicModel}).


There are several aspects that can be seen from Fig.\ref{fig_scan_results}: First, it clearly shows that the lower bound on the $N_1$ mass crucially depends on the lightest active neutrino mass $m_l$. This is a novel realization which, to our knowledge, has not yet been pointed out in the literature. Second, the analytical approximation as stated in Eq.~\eqref{eq_minM1_3RHN} is in very good agreement with the numerical simulation. We are therefore confident that our analytical understanding indeed captures the relevant parameter relations. Third, our analytical approximation is a good approximation for even smaller $m_l$ values than expected (see~Eq.~\eqref{eq_ml_approx_validity}), as we already anticipated at the end Sec.~\ref{subsec_analyticalInsights}. Fourth, even with $\Delta L = 2$ washout being important, $M_1^\text{min}$ still decreases with $m_l$, which one might not necessarily expect.


Although a precise analytical understanding for the region below $m_l \lesssim 10^{-7} \,\text{eV}$, in which $\Delta L = 2$ washout is important, is difficult to come by, we can quantitatively describe what is happening and determine a global lower bound on $M_1^{\text{min}}$. In our case, the way that small values of $m_l$ are achieved is through a suppression of the $N_1$ Yukawa couplings. However, these Yukawa couplings also determine the $N_1$ decay width (see~Eq.~\eqref{eq_decay_width_tree_level}). Smaller values of $m_l$ therefore correspond to a longer $N_1$ lifetime. This, in turn, leads to a partial circumvention of the $\Delta L = 2$ washout because the lepton asymmetry is generated later, explaining why $M_1^{\text{min}}$ decreases with $m_l$.


Nonetheless, there is a limit on how far this is possible and, hence, a limit on $M_1^{\text{min}}$, since the lepton asymmetry generated in $N_1$ decays has to be transformed into a baryon asymmetry by electroweak sphalerons. After sphalerons fall out of equilibrium around $T_\text{sph} \approx 130 \GeV$~\cite{DOnofrio:2014rug}, this conversion is no longer possible. Therefore, we demand that the generation of the baryon asymmetry ends before $z_{B, \,\text{sph}} = M_1 / T_\text{sph}$. Solving the Boltzmann equations with the optimal $\lambda_5$ value for small $m_l$ of $\lambda_5 \sim 10^{-4}$, we find $z_B \approx 3.0 \cdot 10^{-5} (m_l/\text{eV})^{-0.56}$ and a fit to the small $m_l$ region of Fig.~\ref{fig_scan_results} provides $M_1^{\text{min}} \approx 4.6 \cdot 10^7 (m_l/\text{eV})^{0.30}$. Combining both fits with the constraint $z_B < z_{B,\,\text{sph}}$, we first find $m_l \gtrsim 2 \cdot 10^{-12} \,\text{eV}$ and finally
\begin{equation}
M_1^{\text{min}} \sim 10^4 \GeV.
\end{equation}
This is the main result of our paper. Our analysis demonstrates that the scotogenic model with three RHNs allows to realize successful leptogenesis for $N_1$ masses down to $\sim 10 \,\text{TeV}$. The corresponding parameters for which this is possible are given above and in the description of Fig.~\ref{fig_scan_results}. Most notably, a small $N_1$ mass requires an extremely light active neutrino with a mass of $\mathcal{O}\left(10^{-12}\right)\,\textrm{eV}$. The parameter values corresponding to $M_1^{\text{min}} \sim 10^4 \GeV$ determine the neutrino Yukawa matrix via the CI parametrization in Eq.~\eqref{eq_Yukawas_CI_parametrization} together with the PMNS matrix for which we used the best-fit PDG16 values~\cite{Patrignani:2016xqp} with the Majorana phases set to zero. Let us explicitly state this matrix for illustrative purposes, $h=$
\begin{equation}
\label{eq:matrix}
\begin{psmallmatrix}
1 \cdot 10^{-8} + 2 \cdot 10^{-9} \,i & 8 \cdot 10^{-4} + 0 \cdot i & -4 \cdot 10^{-4} + 7 \cdot 10^{-4} \,i \\
1 \cdot 10^{-9} - 6 \cdot 10^{-9} \,i & \phantom{-}9 \cdot 10^{-4} + 7 \cdot 10^{-5} \,i & \phantom{-} 4 \cdot 10^{-3} + 3 \cdot 10^{-15} \, i \\
1 \cdot 10^{-8} - 6 \cdot 10^{-9} \,i & -7 \cdot 10^{-4} + 8 \cdot 10^{-5} \,i & \phantom{-} 4 \cdot 10^{-3} + 1 \cdot 10^{-13} \,i
\end{psmallmatrix} \,.
\end{equation}
This Yukawa matrix looks perfectly natural.%
\footnote{The smallness of some of the imaginary parts is due to our choice of $z_{2 3} = 0$ and not necessary for successful leptogenesis.}
We, thus, conclude that no particular tuning of parameters seems necessary to generate the baryon asymmetry. The only physical assumption we have to make is that the Yukawa couplings of the $N_1$ neutrino must be suppressed compared to those of the $N_{2,3}$ neutrinos, see also footnote~\ref{foot:angles}. Qualitatively, this directly links our findings to  the Sakharov conditions for successful baryogenesis~\cite{Sakharov:1967dj}:~In order to ensure that $N_1$ decays occur sufficiently out-of-equilibrium, we need $h_{1\alpha} \ll 1$, whereas a sufficiently large CP-asymmetry is achieved by sizable $h_{2/3 \,\alpha}$. Similar, qualitative observations have already been made in Ref.~\cite{Hambye:2009pw}. 

Apart from that, the Yukawa couplings in Eq.~\eqref{eq:matrix} have a rather generic and natural structure: the large hierarchies among the different entries are not much larger than those in the charged-lepton Yukawa matrix, and the small absolute values are protected by a global $U(1)$ symmetry which emerges as $h \to 0$. For the same reason, $m_l$ is stable against radiative corrections, too. At this point, it is also important to realize that our optimization in terms of the complex angles in the CI parametrization was not much more than a technical trick. Choosing particular values for the complex angles $z_{12}$, $z_{13}$, and $z_{23}$ must not be considered fine-tuning, as long as the corresponding physical quantities, i.e., the Yukawa couplings $h_{\alpha i}$ do not show any signs of fine-tuning.


The necessary small masses of the lightest active neutrino for low-scale leptogenesis ($m_l \sim 10^{-12} \,\text{eV}$) are an interesting aspect of the model that makes this region of parameter space accessible to experiments. The tritium beta decay experiment KATRIN~\cite{Osipowicz:2001sq}, e.g., intends to perform a direct measurement of the mass of the electron neutrino, $m_{\nu_e}^2\coloneqq \sum_i \left|U_{ei}\right|^2m_i^2$. It will soon start operation and might falsify low-scale leptogenesis in the scotogenic model, if it should find evidence for a lightest active neutrino mass close to its design sensitivity, $m_l \sim 0.2\,\textrm{eV}$. Similarly, the PROJECT~8 collaboration is currently pioneering the development of a next-generation tritium endpoint experiment based on the detection of single-electron cyclotron radiation~\cite{Esfahani:2017dmu}. Once fully developed into a neutrino mass experiment, this approach will allow to probe the entire active neutrino mass range down to $m_l \rightarrow 0$ in the case of an inverted mass hierarchy.

\section{Conclusion}\label{sec_conclusion}

The scotogenic model is the simplest model of radiative neutrino masses and an attractive framework for the unified description of SM neutrino masses, DM and baryogenesis. In this paper, we revisited thermal leptogenesis in the scotogenic model, demonstrating that it allows to accommodate low-scale leptogenesis via the decay of hierarchical RHNs down to rather low RHN masses. In our analysis, we explicitly distinguished between the cases of two and three RHNs. In both cases, we derived a Davidson-Ibarra-type bound on the $CP$ asymmetry. For two RHNs, we argued that leptogenesis inevitably occurs in the strong washout regime. Consequently, this scenario does not substantially differ from standard thermal leptogenesis in the type-I seesaw model. In an explicit analytical calculation, we showed that this follows from the fact that all new prefactors in the scotogenic model essentially cancel. Therefore, for a normal SM neutrino mass hierarchy, the lightest RHN must have a mass of at least $M_1^{\rm min} \sim 10^{10}\,\textrm{GeV}$, while for an inverted mass hierarchy, it must have a mass of at least $M_1^{\rm min} \sim 10^{12}\,\textrm{GeV}$. 


In the 3RHN case, the difference between normal and inverted ordering is negligible and the weak washout regime becomes accessible. However, the efficiency of washout and, hence, $M_1^{\rm min}$ strongly depend on the mass of the lightest active neutrino $m_l$. For $m_l \gtrsim 10^{-7 \,\dotsc\, -6} \,\text{eV}$, the effect of $\Delta L = 2$ washout is negligible and $M_1^{\rm min}$ is directly proportional to $m_l$. For smaller $m_l$ masses, the $\Delta L = 2$ washout becomes important. Nevertheless, even in this regime, the bound on $M_1$ can still be lowered by delaying the decay of the $N_1$ neutrinos. As we were able to demonstrate, this can be achieved by assuming suppressed $N_1$ Yukawa couplings. The generation of the lepton asymmetry is then delayed which allows to circumvent part of the washout. This mechanism is limited by the requirement that leptogenesis must complete before the electroweak sphalerons drop out of equilibrium. In total, we, thus, obtain a global lower limit on the mass of the lightest RHN of $M_1^{\rm min} \sim 10^4\,\textrm{GeV}$. This result needs to be compared with the typical mass bound in standard thermal leptogenesis, $M_1^{\rm min} \sim 10^{9}\,\textrm{GeV}$, which is larger by around five orders of magnitude. We therefore conclude that, in the scotogenic model, one is able to lower the energy scale of leptogenesis compared to the standard type-I seesaw case quite significantly\,---\,without any degeneracy in the RHN mass spectrum!


Another important consequence of the small $N_1$ Yukawa couplings is a very light active neutrino mass eigenstate. In the case of the lowest possible $N_1$ mass, $M_1 \sim 10^4\,\textrm{GeV}$, we find that the lightest active neutrino must have a mass of around $m_l \sim 10^{-12}\,\textrm{eV}$. This is an intriguing prediction that will be tested in future experiments that aim at measuring the absolute neutrino mass scale.


Our analysis in this paper only accounted for the most important effects that are relevant to the generation of the lepton asymmetry: decays and inverse decays of RHNs as well as the associated $\Delta L = 2$ washout processes. In this way, we were able to perform most of our calculations analytically and in a comparatively transparent fashion. In particular, we thus arrived at two important results that are complementary to existing results in the literature: (1) It is, in fact, not possible to realize low-scale leptogenesis in the scotogenic model if one is to work with two RHNs only. (2) On the other hand, in the 3RHN case, it is possible to realize low-scale leptogenesis in the scotogenic model, and in that case, it is not even necessary to assume an approximate RHN mass degeneracy. In future work, it will be interesting to refine the results of our analysis by incorporating several effects that were neglected in this paper. This includes flavor effects, $\Delta L = 1$ scattering processes, and a more careful treatment of kinematic effects in the regime of large $\eta$ masses, $m_\eta \sim M_1$. We expect that such a refined analysis may still lower the absolute lower bound on $M_1$ by a factor of $\mathcal{O}\left(1\cdots10\right)$, similar to standard thermal leptogenesis. Apart from that, we are confident that our qualitative findings will remain unchanged: The scotogenic model is a promising alternative to the type-I seesaw model that allows to lower the energy scale of thermal leptogenesis by many orders of magnitude.

\acknowledgments

The authors thank Jackson D.~Clarke and P.~S.~Bhupal Dev for their valuable contributions at the early stages of this project. The authors are also grateful to Tommi~Alanne, Vedran~Brdar, Marco~Drewes, Thomas~Hambye, Teresa Marrodan Undagoitia, and Stefan Vogl for helpful discussions towards the completion of this project.
M.~P.\ is funded by the International Max Planck Research School for Precision Tests of Fundamental Symmetries (IMPRS-PTFS).
This project has received funding from the European Union's Horizon 2020 research and innovation programme under the Marie Sk\l odowska-Curie grant agreement No.\ 674896 (K.~S.).

\bibliography{literature}
\end{document}